\pdfoutput=1

\documentclass[reprint,twocolumn,amsmath,amssymb,aps,superscriptaddress,prb,eqsecnum]{revtex4-1}
\usepackage{graphicx}
\usepackage{bm}
\usepackage{epsfig}
\usepackage{amssymb}
\usepackage{amsfonts}
\usepackage{amsmath}
\usepackage{color}
\usepackage{xcolor}
\usepackage{epstopdf}
\usepackage{hyperref}
\usepackage{float}
\usepackage{appendix}
\usepackage{wasysym}
\restylefloat{table}
\usepackage{bibentry}
\usepackage{multirow}
\usepackage[caption=false]{subfig}
\usepackage{braket}
\newcommand{\ba}{\begin{eqnarray}}
\usepackage{manfnt}
\newcommand{\ea}{\end{eqnarray}}
\newcommand{\bd}{\begin{displaymath}}

\newcommand{\nn}{\nonumber \\}
\newcommand\scircle{\raisebox{0.25mm}{\scalebox{0.65}{$\CIRCLE$}}}
\newcommand{\RomanNumeralCaps}[1] {\MakeUppercase{\romannumeral #1}}
\DeclareGraphicsExtensions{.pdf,.png,.jpg}
\begin{document}
\title{A New Model for Fractons, Fluxons, and Freeons}

\author{Jintae Kim}
\affiliation{Department of Physics, Sungkyunkwan University, Suwon 16419, Korea}
%\author{Hyun-Yong Lee}
%\email[Electronic address:$~~$]{hyunyong@korea.ac.kr}
%\affiliation{Department of Applied Physics, Graduate School, Korea University, Sejong 30019, Korea}
%\affiliation{Division of Display and Semiconductor Physics, Korea University, Sejong 30019, Korea}
\author{Jung Hoon Han}
\email[Electronic address:$~~$]{hanjemme@gmail.com}
\affiliation{Department of Physics, Sungkyunkwan University, Suwon 16419, Korea}
\date{\today}
\begin{abstract}
We propose a lattice spin model on a cubic lattice that shares many of the properties of the 3D toric code and the X-cube fracton model. The model, made of $\mathbb{Z}_3$ degrees of freedom at the links, has the vertex, the cube, and the plaquette terms. Being a stabilizer code the ground states are exactly solved. With only the vertex and the cube terms present, we show that the ground state degeneracy (GSD) is $3^{L^3 + 3L -1}$ where $L$ is the linear dimension of the cubic lattice. In addition to fractons, there are free vertex excitations we call the freeons. With the addition of the plaquette terms, GSD is vastly reduced to $3^3$, with fracton, fluxon, and freeon excitations, among which only the freeons are deconfined. The model is called the AB model if only the vertex ($A_v$) and the cube ($B_c$) terms are present, and the ABC model if in addition the plaquette terms ($C_{p}$) are included. The AC model consisting of vertex and plaquette terms is the $\mathbb{Z}_3$ 3D toric code. The extensive GSD of the AB model derives from the existence of both local and non-local logical operators that connect different ground states. The latter operators are identical to the logical operators of the $\mathbb{Z}_3$ X-cube model. Fracton excitations are immobile and accompanied by the creation of fluxons - plaquettes having nonzero flux. In the ABC model, such fluxon creation costs energy and ends up confining the fractons. Unlike past models of fractons, vertex excitations are free to move in any direction and picks up a non-trivial statistical phase when passing through a fluxon or a fracton cluster.  
\end{abstract}
\maketitle

\section{Introduction}
We have witnessed significant convergence of ideas from quantum information, quantum computation, and quantum many-body theory in the past decades~\cite{wen19}. Since Shor introduced the idea of error correction~\cite{shor95}, a large number of papers have followed suit~\cite{laflamme97, kitaev03, gottesman97, gottesman96, shor97, shor96, kitaev98, freedman01, steane96, zurek98, werner01} firmly establishing the notion of quantum computation and quantum memory. As a model for quantum memory, the stability of quantum bits stored in the toric code~\cite{kitaev03} or its annular~\cite{kitaev98, freedman01} and higher dimensional versions~\cite{simon13, horodecki10, wang12} derives from having to invoke a non-local string operator to move one ground state into a different one. The idea of self-correcting quantum memory also has been proposed and improved~\cite{bacon06, haah13, michnicki14}. Nowadays, the confluence is especially apparent in the works on fractons~\cite{hermele19, vijay16, haah11,  vijay17, slagle17, slagle18, chen18, chamon05, vijay15}.

Fracton models are either gapped or gapless. Further, gapped fracton models can be classified as type \RomanNumeralCaps{1} or type \RomanNumeralCaps{2}~\cite{hermele19}. Type \RomanNumeralCaps{1} gapped fracton models have two distinct quasi-particles: fractons which are immobile, and sub-dimensional particles such as lineons and planons which can move along a line or within a plane, respectively. The X-cube model~\cite{vijay16} is a simple realization of the type \RomanNumeralCaps{1} gapped fracton models and a natural generalization of the toric code to three dimensions. Type \RomanNumeralCaps{2} gapped fracton models can have only fractons as quasi-particles and the well-known example is Haah's cubic code~\cite{haah11}. The number of quantum bits becomes equivalent to the ground state degeneracy (GSD) in topological models of quantum memory and grows sub-extensively, i.e. linear in the system size $L$, for the gapped fracton models proposed so far~\cite{hermele19}. Gapless fracton models can be understood to some extent in the framework of U(1) symmetric tensor gauge theory~\cite{pretko17,pretko172} which describes well the origin of the restricted mobility of sub-dimensional particles.

In this paper, we propose a different kind of fracton model. Written in terms of $\mathbb{Z}_3$ degrees of freedom on the links of the $L\times L\times L$ cubic lattice, our model consists of mutually commuting vertex, cube, and plaquette operators. When only the vertex and the cube operators are present, the model supports extensive GSD equal to $3^{L^3+3L-1}$. With the addition of the plaquette term, GSD becomes $3^3$. The $\mathbb{Z}_N$ generalization of the $X$-cube model suggested previously~\cite{vijay17, slagle17, slagle18, chen18} predicts, on the other hand, a sub-extensive GSD: $\log_{N} {\rm GSD} \sim O(L)$. The vast increase in the GSD is understood in terms of local symmetries in our model, absent in previous fracton models. Some of the ground states are connected by a local operation (therefore not topological) but some are only connected by non-local loop operator as in other models of topological quantum computation. We identify such local as well as non-local operators in the model.

In addition to fracton excitations that are immobile as usual, there are vertex excitations in our model whose motion is ``free", unlike in other fracton models predicting one-dimensional confinement of the vertex excitation. Our vertex excitations are thus called {\it freeons} in contrast to {\it lineons} in previous fracton models. The third kind of excitation supported in our model is the {\it fluxon}, which is an analogue of $m$ particles in the toric code. Unlike the toric code, however, these fluxons cannot exist in isolation but must always share an edge with another fluxon and therefore always form a cluster. The boundary of such cluster then forms a flux loop. A freeon passing through such a loop picks up an Aharonov-Bohm phase, in a generalization of the statistical phase factors picked up by $e$ and $m$ particles in the $\mathbb{Z}_N$ toric code. In essence, the $m$ particles of the toric codes are replaced by ``$m$ tubes'', inside which the flux is confined. The adiabatic motion of a freeon can be used also as a detection scheme for fractons, as described in other works~\cite{hermele18, chen19, slagle17, iadecola19, hermele19a}. 
%can interact with a fracton as a sub-dimensional particle in fracton models and interact with a fluxon as a quasi-particle of vertex excitation in the 3D toric code.

Section \ref{sec:review} makes a self-consistent review of the toric code and the X-cube models in a language and notation that will be consistently used in the remainder of the paper. The model we propose is introduced in Sec. \ref{sec:ABC}. Counting of the ground state degeneracy is performed carefully in Sec. \ref{sec:local-symmetry} followed by the analysis of logical operators in Sec. \ref{sec:logical-operator}. The fracton, fluxon, and freeon excitations of the model are defined and their characteristics and statistical interactions analyzed in Sec. \ref{sec:excitation}. 

\section{Review of $\mathbb{Z}_2$ toric code \\and X-cube model}
\label{sec:review}

The mathematical structures of the toric code and the X-cube model are well known. The toric code has $\mathbb{Z}_2$ degrees of freedom residing on the links of a square lattice. There are vertex operators $A_v$ and plaquette operators $B_p$ defined respectively as
\ba
A_v & = & \frac{1}{2} (1+ \prod_{i\in +_v} x_i)  \nn
B_p & = & \frac{1}{2} (1+ \prod_{j\in \square_p} z_j)  \label{eq:A-and-B} \ea
in terms of the Pauli operators $x$ and  $z$ at the links. The subscript $i$ refers to the four links emanating from a vertex $v$, and $j$ to the four links surrounding a given plaquette $p$.
Both operators are projectors $A_v^2 = A_v$, $B_p^2 = B_p$ and define the toric code Hamiltonian
\ba H = -\sum_v A_v - \sum_p B_p  \ea
as the sum over all the vertices and the links of the lattice. The ground state(s) of the model is found by either projecting the ground state of $-\sum_v A_v$ with $\prod_p B_p$, or by projecting the ground state of $-\sum_p B_p$ with $\prod_v A_v$. In each case, we obtain the ground states
\ba |G_1 \rangle &=& [ \prod_p B_p ] \ket{S_A} \text{ or }\nn
|G'_1 \rangle &=& [ \prod_v A_v ] \ket{S_B} \label{eq:GandG'} \ea
where the seed states $\ket{S_A}$ and $\ket{S_B}$ respectively satisfy $A_v\ket{S_A}=\ket{S_A}$ and $B_p\ket{S_B}=\ket{S_B}$ for arbitrary $v$ and $p$. One simple example of $\ket{S_A}$ and $\ket{S_B}$ are given as a product of $|0\rangle_l$'s and $|\overline{0}\rangle_l$'s over all the links $l$, which are the eigenstates of $z$ and $x$ operators with $z_i|0\rangle_i = |0\rangle_i$ and $x_i|\overline{0} \rangle_i = |\overline{0} \rangle_i$, respectively. In general, we have $|G_1 \rangle \neq |G'_1 \rangle$ as they possess different sets of quantum numbers, to be clarified below.

\begin{figure}[h]
\centering
\includegraphics[width=0.40\textwidth]{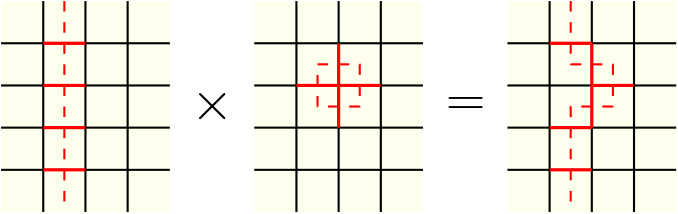} 
\caption{A deformed path in the toric code is obtained by acting an elementary plaquette operator $B_p$ to an existing path. The new path defines the same logical operator as the old one.}
\label{fig:deformed-path}
\end{figure}

The toric code possesses string operators that commute with the Hamiltonian:
\ba Z^h = \prod_{i \in h_l} z_{i} , ~~~  Z^v = \prod_{i \in v_l} z_{i} \label{eq:Zh-and-Zv} . \ea
The product runs over a horizontal line labeled $h_l$, or a vertical line labeled $v_l$, in the square lattice. There is a second pair of string operators given by
\ba X^h = \prod_{i \in h_d} x_i  , ~~~ X^v = \prod_{i \in v_d} x_i . \label{eq:Xh-and-Xv}\ea 
This time the strings pass through the dual lattice points at the center of the plaquettes, as shown in Fig. \ref{fig:deformed-path}. The two sets of string operators obey the algebra
\ba X^h Z^v = - Z^v X^h ~~~X^v Z^h = - Z^h X^v .  \ea

On a torus, new ground states are generated from $|G_1\rangle$ in Eq. (\ref{eq:GandG'}) through the action of string operators
\ba |G_2\rangle & = & X^h |G_1 \rangle , 
%= \prod_v A_v (\prod_{i \neq i'} |0\rangle_i \prod_{i' \in h_d} |1 \rangle_{i'} ) , 
\nn
|G_3 \rangle & = & X^v  |G_1 \rangle , \nn
|G_4 \rangle & = & X^v X^h |G_1 \rangle . \ea 
It turns out the other pair of string operators characterizes the four ground states as 
\ba Z^h |G_1 \rangle  = +  | G_1 \rangle ,  & ~~ & Z^v |G_1 \rangle  =  + | G_1 \rangle \nn
Z^h |G_2 \rangle =  + | G_2 \rangle ,  & ~~ & Z^v |G_2 \rangle =  - | G_2 \rangle \nn
Z^h |G_3 \rangle =  -  | G_3 \rangle , & ~~ & Z^v |G_3 \rangle =    + | G_3 \rangle \nn
Z^h |G_4 \rangle =  -  | G_4 \rangle , & ~~ & Z^v |G_4 \rangle =  -  | G_4 \rangle .
\ea 
Each ground state is labeled with a pair of binary quantum numbers corresponding to the eigenvalues of $(Z^h , Z^v )$. Alternatively, one can choose the four ground states as $|G'_1 \rangle$ and 
\ba 
%|G'_1 \rangle & = & [ \prod_p B_p ] (\prod_i |\overline{0}\rangle_i ) \nn
%
|G'_2 \rangle & = & Z^h |G'_1 \rangle  
%= [ \prod_p B_p ] (\prod_{i \neq i'} |\overline{0} \rangle_{l} \prod_{i' \in h_l} |\overline{1}\rangle_{l'} 
\nn
|G'_3 \rangle & = & Z^v |G'_1 \rangle   \nn
|G'_4 \rangle & = & Z^v Z^h |G'_1 \rangle . \ea
These four states are in turn labeled by 
\ba X^h |G'_1 \rangle  = +  | G'_1 \rangle ,  & ~~ & X^v |G'_1 \rangle  =  + | G'_1 \rangle \nn
X^h |G'_2 \rangle =  + | G'_2 \rangle ,  & ~~ & X^v |G'_2 \rangle =  - | G'_2 \rangle \nn
X^h |G'_3 \rangle =  -  | G'_3 \rangle , & ~~ & X^v |G'_3 \rangle =    + | G'_3 \rangle \nn
X^h |G'_4 \rangle =  -  | G'_4 \rangle , & ~~ & X^v |G'_4 \rangle =  -  | G'_4 \rangle . 
\ea 
The string operators $X, Z$ are also known as logical operators for their implication in quantum information~\cite{kitaev03} - a term we continue to adopt in the rest of the paper.  
%From now on, we adopt the $z$-basis where $X_h$ and $X_v$ serve as the logical operators. The action of $X_h$, for instance, flips the seed spins $|0\rangle \rightarrow |1\rangle$ along a horizontal row of links. 
%We will also adopt the nomenclature $H \equiv X_h$ and $V \equiv X_v$ for the logical operators extending along the horizontal and the vertical directions, respectively.

The X-cube model is a generalization of the toric code to three dimensions. There are three kinds of vertex operators per vertex $v$:
\ba
A_v^{xy} & = & \frac{1}{2} (1+ a_v^{xy}) , ~~ a^{xy}_v = \prod_{i\in +_{v, xy} } x_i  \nn
A_v^{yz} & = & \frac{1}{2} (1+ a_v^{yz}) , ~~ a^{yz}_v = \prod_{i\in +_{v, yz} } x_i  \nn
A_v^{xz} & = & \frac{1}{2} (1+ a^{xz}_v ) , ~~ a^{xz}_v = \prod_{i\in +_{v, xz} } x_i . \label{eq:X-cube-A} 
\ea
Each one has the same form as the vertex operator of the toric code, but now there are three planes $xy$, $yz$, and $xz$ in which to define them. The symbol $i\in +_{v, xy}$ for instance means the four links emanating from the vertex $v$ in the $xy$ plane. Instead of the plaquette operator $B_p$ in the toric code, one has the cube operator $B_c$:
\ba 
~~~B_c=\frac{1}{2} (1+ b_c ) , ~~ b_c = \prod_{j\in \square_c} z_j  .  \label{eq:Bc} \ea
There are twelve $z$ operators coming from the edges of a cube $c$. All four operators $ A^{xy}_v,~ A^{yz}_v,~ A^{xz}_v,~ B_c $ are projectors and mutually commuting. The GSD of this model is known to be $2^{6L-3}$ for a $L\times L\times L$ cubic lattice under the periodic boundary conditions (PBC) in all three directions (a three-torus)~\cite{vijay16,hermele17,chen18}. The factor $6L-3$ in the exponent is indicative of the number of independent logical operators in the X-cube model. 

Firstly a ground state of the X-cube model is found by the projection
\ba &&|G\rangle = [ \prod_v A^{xy}_v A^{yz}_v A^{xz}_v ]  ( \prod_l |0\rangle_l )  ~~~ {\rm or} \nn
&&|G' \rangle = [ \prod_c B_c ] ( \prod_l |\overline{0} \rangle_l ) \label{eq:X-cube-GS} \ea
in close analogy to the ground state construction of the toric code. The logical operators are exactly those of the toric code, Eqs. (\ref{eq:Zh-and-Zv}) and (\ref{eq:Xh-and-Xv}), but now they exist for each planar orientation $xy$, $yz$, and $xz$. Application of one of these logical operators on the ground state $|G\rangle$ shown in Eq. (\ref{eq:X-cube-GS}) brings it to another ground state of the X-cube Hamiltonian. 

We make a careful discussion of the translational invariance property of the logical operators in the toric code or the X-cube model. The horizontal string operator, for instance, is defined for an arbitrary vertical position of the square lattice and vice versa and yet, exactly the same state results from their actions on the ground state irrespective of their vertical positions. The reason for this is a special property of the link operators, 
\ba
A_v^{\alpha} & = & ( \prod_{i\in +_{v, \alpha} } x_i )  A_v^{\alpha} \label{eq:Av-identities}
\ea
where $\alpha = xy,~ yz,~ xz$. It states that any vertex operator $A_v^{\alpha}$ can be interpreted equally well as an additional operation $\prod_{i\in +_{v, \alpha} } x_i$ followed by $A_v^{\alpha}$ operation itself. What the extra operation $\prod_{i\in +_{v, \alpha} } x_i$ does is to flip the seed spin states $|0\rangle_l$ on the four links tied to a vertex $v$ in the $
\alpha$ plane: $|0\rangle_l \rightarrow |1\rangle_l$. Due to the identity mentioned in Eq. (\ref{eq:Av-identities}), one might as well absorb $\prod_{i\in +_{v, xy} } x_i$ as part of the logical operator,
\ba X^{xy}_v A^{xy}_v = ( X^{xy}_v \prod_{i\in +_{v, xy} } x_i ) A_v^{xy}. \label{eq:X-identity} \ea
The upper index in $X^{xy}_v$ means that it is the vertical logical operator within the $xy$ plane of the cubic lattice. The new logical operator $X^{xy}_v \prod_{i\in +_{v, xy} } x_i$ has a trajectory that is bent by one elementary square unit, in a manner depicted in Fig. \ref{fig:deformed-path}, but its action on a ground state $|G\rangle$ results in exactly the same state as before due to identities mentioned in Eqs. (\ref{eq:Av-identities}) and (\ref{eq:X-identity}). In particular, a shift of the entire vertical trajectory of $X^{xy}_v$ by one lattice spacing along the horizontal direction results in the same logical operator. This is why there are only two independent logical operators, one horizontal and one vertical, in the toric code. 

There are two logical operators per plane, per layer, per orientation (i.e. $xy$, $yz$, $xz$), which make up the $6L$ logical operators overall in the X-cube model. Certain constraints exist among these logical operators~\cite{hermele17}, and reduce the number of independent operators from $6L$ to $6L-3$. 
This, in turn, explains ${\rm GSD} = 2^{6L-3}$ of the X-cube model~\cite{vijay16, hermele17, chen18}. 

We present another way to obtain the same GSD. A general theory of GSD for the stabilizer codes as worked out in Refs.~\onlinecite{gottesman96,shor97} says
\ba \log_N {\rm GSD} = N_l - N_{s} = N_{lo} . \label{eq:GSD-formula} \ea
Each symbol represents the degrees of freedom ($N$) residing at the link, the number of links ($N_l$),  of independent stabilizers ($N_s$), and of independent logical operators ($N_{lo}$), respectively. For instance, in the toric code, the number of independent stabilizers is $N_s = 2L^2 -2$ because among the $2L^2$ stabilizers there exist two constraints. One can obtain $N_{lo} =2$ alternatively from the argument on the number of independent logical operators presented above. Previous counting argument for the GSD of the X-cube model~\cite{hermele17,chen18} focused on $N_{lo}$. Below we show how to count $N_{s}$ and arrive at the same GSD. 

Ostensibly, there are $4L^3$ stabilizers in the X-cube model far exceeding even the number of links, $3L^3$. One quickly notices though that $a_v^{xz} = a_v^{xy} a_v^{yz}$ at every vertex $v$, allowing for only $2L^3$ independent vertex stabilizers. Furthermore, there are certain identities obeyed among the vertex operators
\ba \prod_{v \in i-{\rm th} ~ \alpha ~ {\rm plane} }  a_v^{\alpha} = 1  \label{eq:2.16} \ea
where $\alpha = xy,~ yz,~ xz$. The product of $a_v^{xy}$ operators over the vertices in a given $xy$ plane is an identity and so on. Altogether there are $3L$ such identities among the vertex stabilizers, which seems to reduce the number of independent vertex stabilizers from $2L^3$ to $2L^3 - 3L$. Such counting is still incomplete, as there is yet another identity 
\ba & & \left( \prod_{i=1}^L  \prod_{v \in i-{\rm th} ~ xy ~ {\rm plane} }  a_v^{xy}   \right) \left( \prod_{i=1}^L  \prod_{v \in i-{\rm th} ~ yz ~ {\rm plane} }  a_v^{yz} \right)   \nn
& &  ~~~~~~~
= \left(  \prod_{i=1}^L  \prod_{v \in i-{\rm th} ~ xz ~ {\rm plane} }  a_v^{xz}  \right)   \label{eq:2.17} \ea
which follows readily from $a^{xy}_v a^{yz}_v a^{xz}_v = 1$. Instead of having $3L$ independent identities shown in Eq. (\ref{eq:2.16}), we have only $3L-1$ identities due to Eq. (\ref{eq:2.17}) hence the total number of independent vertex operators is $2L^3 - 3L +1$. There are analogous constraints among the cube operators $b_c$,
\ba \prod_{c \in i-{\rm th} ~  {\alpha} ~ {\rm layer} } b_c & = &  1 , \label{eq:2.18} \ea
where $\alpha = xy,~ yz,~ xz$ layer refers to a single stack of cubes parallel to the $\alpha$ plane. There are $3L$ identities overall among the cube operators but here we also find some extra relations among the identities
\ba  & & \prod_{i=1}^L \prod_{c \in i-{\rm th} ~  {xy} ~ {\rm layer} } b_c   \nn
& = & \prod_{i=1}^L \prod_{c \in i-{\rm th} ~  {yz } ~ {\rm layer} } b_c  \nn
& = & \prod_{i=1}^L \prod_{c \in i-{\rm th} ~  {xz} ~ {\rm layer} } b_c . \label{eq:2.19} \ea
This gives two extra relations among the identities mentioned in Eq. (\ref{eq:2.18}) and reduces the number of constraints from $3L$ to $3L-2$. As a result, the number of independent cube operators is $L^3 - 3L +2$. The number of independent stabilizers is therefore
\ba N_s &=& ( 2L^3 - 3 L +1 ) + (L^3 - 3L +2) \nn
&=& 3L^3 - 6L +3 \ea
which directly leads to the well-known result ${\rm GSD} = 2^{N_l - N_s} = 2^{6L -3}$ for the X-cube model. This way of counting the number of independent stabilizers seems cumbersome in the case of the X-cube model but will prove valuable when it comes to calculating GSD of our model. 

\section{The ABC model}
\label{sec:ABC}

\begin{figure}[h]
\centering
\includegraphics[width=0.42\textwidth]{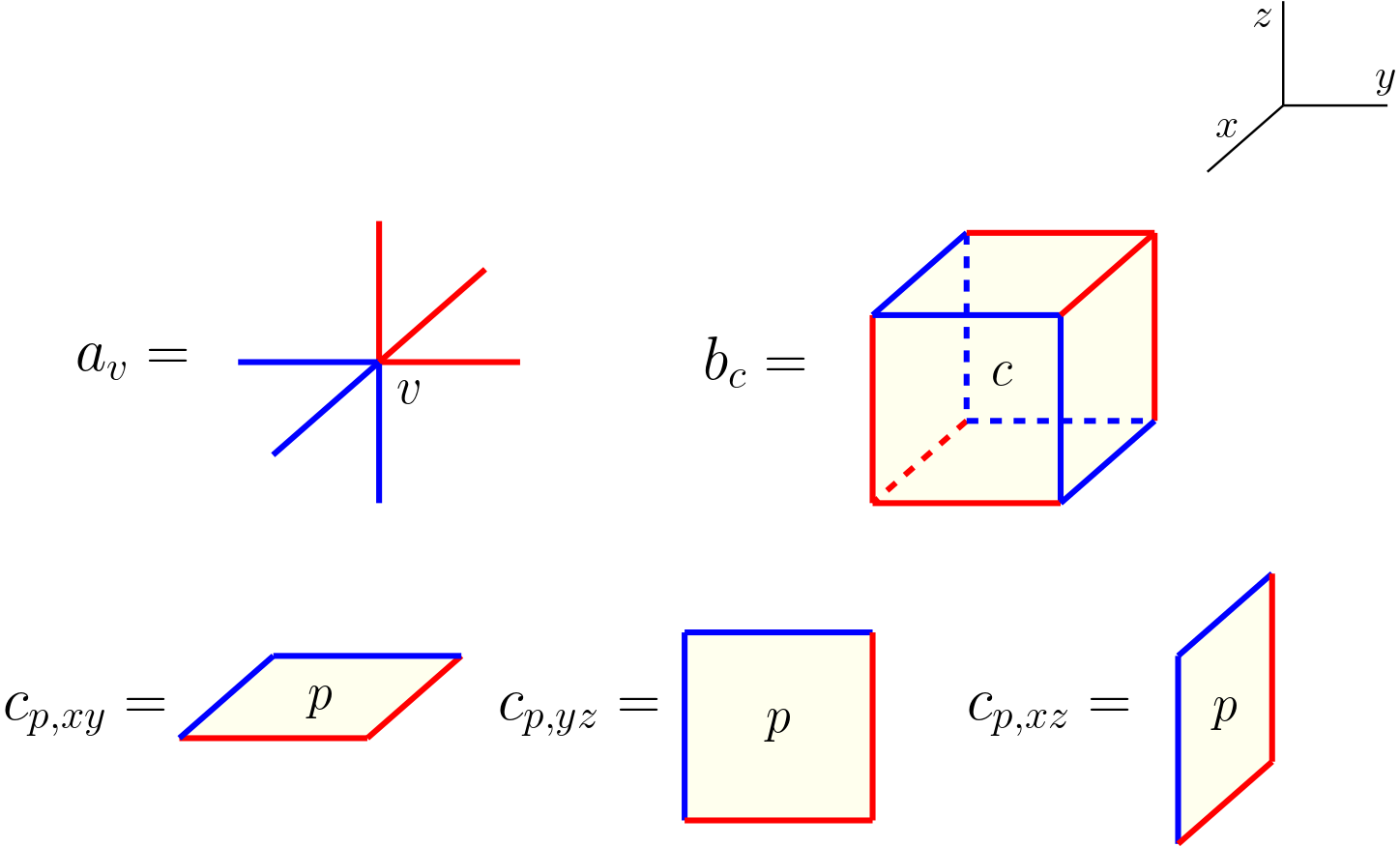} 
\caption{Definitions of vertex, cube, and plaquette operators in our model. Red and blue lines respectively represent $x$ and $x^2$ for $a_v$, and $z$ and $z^2$ for $b_c$ and $c_{p,\lambda}$'s. The $x,~y,~z$ orientations shown here will be adopted for all subsequent figures.}
\label{fig:2}
\end{figure}

The model we propose consists of the vertex ($A_v$), the cube ($B_c$), and the plaquette $(C_p)$ terms: 
\ba
H&=&-\sum_v A_v-\sum_c B_c-\alpha \sum_{p} C_p .
\ea
Like its predecessors, it is a stabilizer code on a cubic lattice. The link variables are specifically chosen to be $\mathbb{Z}_3$. We continue to adopt notations $x$ and $z$ for the operators which obey the relations ($|g\rangle = |0\rangle,~ |1\rangle,~ |2\rangle$)
\ba
&& x\ket{g}=\ket{g+1} (\text{mod}~ 3)~~~~ z\ket{g}=\omega^g \ket{g} . \label{eq:d3 relation} \ea
As a result we get $zx=\omega xz$ where $\omega=e^{2\pi i/3}$. We define the vertex and the cube operators by
\ba A_v & = & \frac{1}{3}(1+a_v+a_v^2) \nn
B_c & = & \frac{1}{3} (1+b_c+b_c^2) . \ea
Here, $a_v$ is the product of $x$'s and $x^2$'s defined on the six links connected to a vertex as shown in Fig. \ref{fig:2}. The cube operator $b_c$ is the product of $z$'s and $z^2$'s defined the twelve edges of a cube as illustrated in Fig. \ref{fig:2}. In addition, there is a plaquette operator $C_{p, \lambda}$ ($\lambda=xy,~yz,~xz$) for each plaquette of the lattice,
\ba
C_{p, \lambda}&=&\frac{1}{3}(1+c_{p, \lambda}+c_{p, \lambda}^2),
\ea
with $c_{p,\lambda}$ defined as in Fig. \ref{fig:2}. The $x,~y,~z$ directions of the cubic lattice shown in Fig. \ref{fig:2} will be adopted in all subsequent figures. All the operators $A_v,~ B_c,~ C_{p, \lambda}$ are commuting. Our $\mathbb{Z}_3$ model is distinctly different from the $\mathbb{Z}_N$ ($N=3$) generalizations of the X-cube model given previously~\cite{vijay17, slagle17, slagle18, chen18}. %A striking feature of the AB model ($\alpha=0$) is its extensive GSD: ${\rm GSD} = 3^{L^3 + 3L-1}$. For the full ABC model ($\alpha >0$), we have ${\rm GSD}=3^3$.

Computing the GSD of the AB model requires that we work out either $N_s$ or $N_{lo}$ in the general formula, Eq. (\ref{eq:GSD-formula}). The number of independent stabilizers $N_s$ is easy to work out. There are $L^3$ vertex stabilizers in the AB model, but only $L^3-1$ of them are independent due to the identity $\prod_v a_v=1$ (product over all the vertices of the cubic lattice). For the cube operators, the constraints given in Eqs. (\ref{eq:2.18}) and (\ref{eq:2.19}) apply to the AB model as well, with the appropriate modification of the definition of the $b_c$. As in the X-cube model, only $L^3-3L+2$ cube stabilizers are independent. The overall number of independent cube and vertex stabilizers is
\ba
N_s &=&( L^3-1 ) + ( L^3-3L+2 ) \nn
&=&2L^3-3L+1,
\ea
meaning that the GSD of the AB model is
\ba \log_3 ( {\rm GSD} ) = L^3+3L-1 ~~ ({\rm AB ~ model}) \label{eq:GSD-AB} \ea  
The GSD here is {\it extensive}. What happens to the GSD for the full ABC model requires an understanding of the local plaquette symmetry and the notion of independent $p$-sectors, which is discussed in the next section. Once this is cleared, GSD of the ABC model follows as $3^3$. 

\section{Local plaquette symmetry}
\label{sec:local-symmetry}

Eigenstates of the ABC model can be labeled in terms of the eigenvalues of the plaquette operators $c_{p, \lambda}$ as they commute with the rest of the terms in the model.  The $3^{3L^3}$-dimensional Hilbert space divides up into various sectors according to the eigenvalues of the plaquette operators taking on $1,~ \omega$, or $\omega^2$. Each ``$p$-sector'' is then characterized by a set of $3L^3$ values $\{ p_i \}$ ($p_i = 1,~ \omega,~ \omega^2$) where $i$ refers to a plaquette. At first, there seems to be $3^{3L^3}$ distinct $p$-sectors, but more careful reasoning says otherwise. 

As a warm-up, we consider the situation of a two-dimensional square lattice where each elementary square plaquette carries one of the three eigenvalues of the $c_{p, xy}$ operator. The Hilbert space dimension for a $L\times L$ square lattice is $3^{2L^2}$. There are $L^2$ plaquette operators in total, but the product of all the plaquette operators in the square lattice equals 1. The number of independent $p$-sectors becomes $3^{L^2 -1}$ instead of $3^{L^2}$, also implying that within each $p$-sector the dimension of the Hilbert space must be $3^{L^2 +1}$. One can in fact check that this is so, by explicit counting of the number of distinct link configurations $\{ z_i \}$ ($z_i = 0,~ 1,~ 2$) that are consistent with a given distribution of plaquette quantum numbers. Take, for instance, the case where $p_i = 1$ (zero flux) for all the elementary plaquettes, requiring the four-link variables to obey the condition $z_1 + z_2 = z_3 + z_4 $ (mod 3). Careful counting of all the possible $z$'s consistent with the constraints indeed yields the desired result $3^{L^2 + 1}$. The $3^{2L^2}$-dimensional Hilbert space factorizes as 
\ba 3^{2L^2} = 3^{L^2 -1} \times 3^{L^2 +1}  . \ea
A similar consideration gives the factorization of the Hilbert space of the ABC model as 
\ba 3^{3L^3}= 3^{2L^3 - 2} \times 3^{L^3 +2}  . \label{eq:3d link}\ea
The number of independent $p$-sectors is $3^{2L^3 -2}$ while the number of states in a given $p$-sector is $3^{L^3 + 2}$. 

The gist of the counting argument in both two and three dimensions can be explained. One starts with a single, two-dimensional square lattice. By explicit counting, one can prove that the number of independent link configurations, consistent with the constraint, in the first row of squares is $3^{2L}$. For the subsequent rows, the number of independent link configurations is reduced to $3^L$ per row, except the last row where only 3 possible link values are allowed. In total, the number of allowed link configurations in the two-dimensional square lattice, under the PBC, is
\ba 3^{2L} \times 3^L \times \cdots 3^L \times 3 = 3^{L^2 + 1} . \label{eq:2D link}\ea
\begin{figure}[h]
\centering
\includegraphics[width=0.40\textwidth]{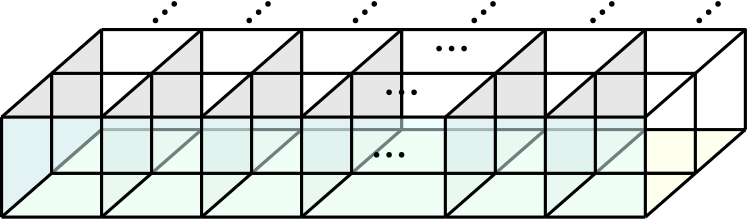} 
\caption{The first layer of the cubic lattice. The plaquettes that contribute to the counting of the independent link configurations are colored. Floor plaquettes, front plaquettes, side plaquettes are colored in yellow, blue, grey, respectively.}
\label{fig:8}
\end{figure}

The counting argument for the number of link configurations in three dimensions proceeds similarly, by starting with the first layer of cubes (Fig. \ref{fig:8}). Counting the link configurations of the bottom $xy$ plane (yellow) is already done and gives $3^{L^2 +1}$. The numbers of link configurations of the front plaquettes (blue, facing us) and the side plaquettes (grey, at right angles to the front plaquettes) are $3^L$ and $3^{L(L-1)}$, respectively, by explicit counting. The rest of the plaquettes at the back and the top carry no further degrees of freedom in the link variables. For the second to the $(L-1)$-th layers, the number of link configurations for the front and the side plaquettes are the same as in the first layer, but the floor plaquettes no longer need to be counted because their configurations have been fixed from the layer below. For the final, $L$-th layer, there are only three link configurations allowed after taking into account the PBC. Tallying the count, we get 
\ba 3^{L^2+1} \times (3^L3^{L(L-1)})^{L-1}\times 3 =3^{L^3+2}
\ea
for the number of link configurations in a given $p$-sector. This explains the factor $3^{L^3 +2}$ in the factorization, Eq. (\ref{eq:3d link}). 

\begin{figure}[h]
\centering
\includegraphics[width=0.47\textwidth]{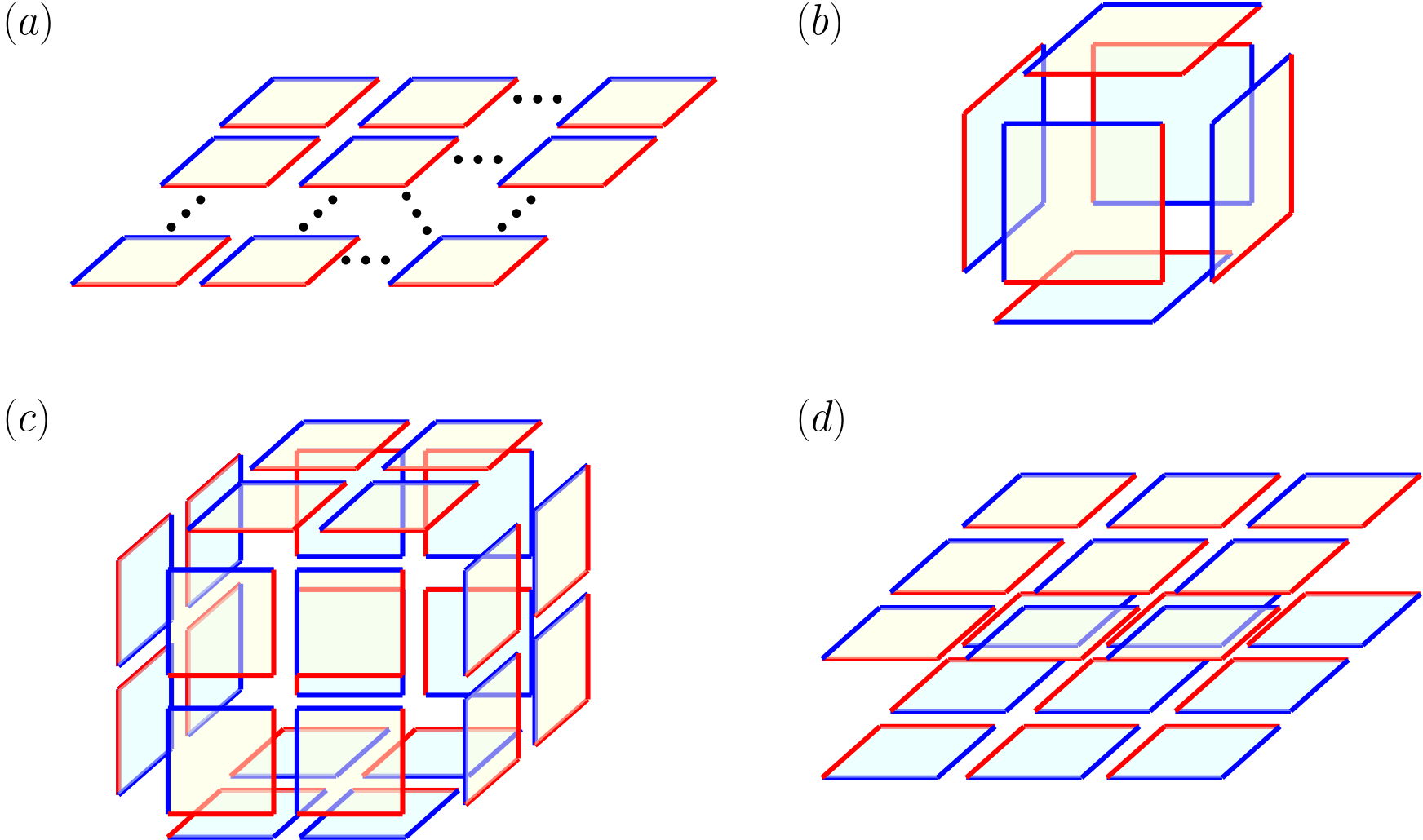} 
\caption{Various constraints on the plaquette operators. (a) Product of $c_{p}$'s in a given $xy$ layer has to be $1$. The same condition applies for $yz$ and $xz$ layers as well. (b) Product of three $c_p$'s (yellow) and three $c_p^2$'s (blue) of a cube in this manner gives $1$. (c) The product of all cube operator $c_p$'s results in surface contributions only, which then becomes 1 under the PBC. (d) The product of cube operators in a single layer is equivalent to a product of plaquette operators at the top and the bottom.}
\label{fig:7}
\end{figure}

Next, although this is not strictly necessary, we count the number of independent $p$-sectors. The correct answer must be $3^{3L^3} / 3^{L^3 +2} = 3^{2L^3 -2}$, but an independent check will be highly desirable. In the case of two-dimensional lattice, we saw that one of the plaquette numbers is fixed entirely in terms of the remaining $L^2 -1$ plaquette numbers. In the cubic lattice, there are certain constraints associated with each cube. Note that the product of plaquette operators on the six faces of a cube in the manner depicted in Fig. \ref{fig:7}(b) equals one. The plaquette numbers must satisfy a similar constraint that their product is equal to one for each cube. There are $L^3$ cubes but only $L^3 -1$ cube constraints since the product of all cube constraints automatically gives 1 and one of the constraints can be expressed as the product of the remaining $L^3 -1$ cube constraints. 

We are still not completely done. In the case of a two-dimensional lattice, there are planar constraints like the one in Fig. \ref{fig:7}(a), and even in a three-dimensional lattice, these constraints must still hold. However, the planar constraints are not entirely independent from the cube constraints mentioned in the previous paragraph. To see why, take the product of all the cube constraints in one layer, as shown in Fig. \ref{fig:7}(d). It is easy to check that the operators on the side of the layer become one (hence not shown in Fig. \ref{fig:7}(d)), while the top and the bottom faces of the layer give the product of $p_i$'s. From the condition that the product of all cube constraints in a layer is one, we infer the following relation among the product of $p_i$'s in the two planes: 
\ba
\prod_i p_{xy,i}=\prod_{j} p_{xy,j} . \label{eq:plane-to-plane}
\ea
The indices $i$ and $j$ refer to the plaquettes of the upper and the lower plane, respectively. As a result, there are planar constraints but for only one of the planes in a given orientation. One may think of Eq. (\ref{eq:plane-to-plane}) as ``constraints among constraints'', so to speak. With three such constraints, one per orientation of the planes, we finally come to the number of independent constraints on the plaquette operators $L^3 -1 +3 = L^3 +2$. The number of distinct $p$-sectors is then $3^{3L^3 - ( L^3 + 2 ) } = 3^{2L^3 -2}$ as desired.

\begin{figure}[h]
\centering
\includegraphics[width=0.30\textwidth]{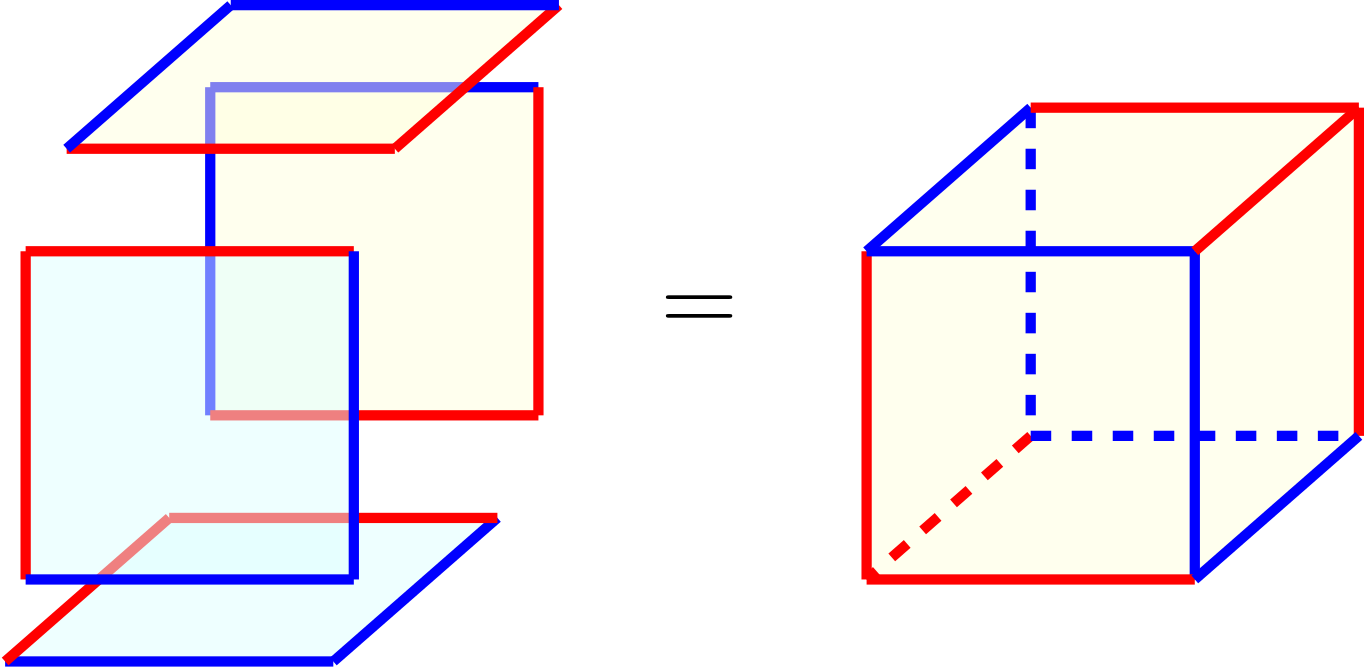} 
\caption{The product of $c_{p}$ ($c_{p}^2$) on upper (lower) $xy$ plane of a cube and $c_{p}$ ($c_{p}^2$) on the front (back) $yz$ plane of a cube is equal to the cube operator $b_c$.} 
\label{fig:pc}
\end{figure}

Finally, we come to the task of calculating GSD of the full ABC model. It is first of all essential to realize that the cube stabilizer is no longer an independent operator, in the sense that it can be decomposed as a product of four plaquette operators as shown in Fig. \ref{fig:pc}. The cube operators must be ruled out in the counting of the number of independent stabilizers. On the other hand, the number of independent plaquette operators is derived straightforwardly from the number of independent $p$-configurations, which we worked out to be $3^{2L^3 - 2}$ earlier. The number of independent $A_v$-stabilizers is $L^3 -1$ as mentioned earlier. Overall, we get the number of independent stabilizers and the GSD in the ABC model as
\ba N_s &=& (L^3 -1 ) + (2L^3 - 2) = 3L^3 - 3 \nn
{\rm GSD} & = & 3^{N_l - N_s} = 3^3 ~~ ({\rm ABC ~ model} ) . \ea
As far as counting of the GSD goes, the AC model (without the cube term) is the same as the full ABC model.

\begin{figure}[h]
\centering
\includegraphics[width=0.40\textwidth]{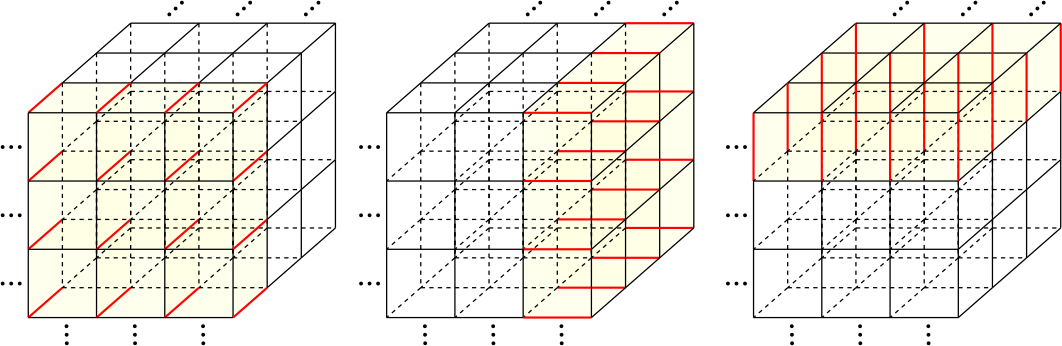} 
\caption{Three independent membrane operators generating the ground states in the ABC model.} 
\label{fig:mo}
\end{figure}

The ground states of the ABC model arise in the $p_i = 1$ sector (zero flux for all the plaquettes). There are only three logical operators $N_{lo}=3$ generating the ground state manifold. When we start from a ground state $[\prod_v A_v] (\otimes_l \ket{0}_l)$, logical operators that connect this ground state to other ground states are shown in Fig. \ref{fig:mo} as the product of $x$'s occupying an entire ``membrane'' and are distinguished from the string operators of the toric code or the X-cube model. Translating a membrane operator by one lattice constant is an identity operation, in the same sense that the translation of the logical operator in the toric code or the X-cube is an identity. There is one membrane logical operator per plane orientation and three overall. 

Identifying the logical operators generating the ground states of the AB model is much more challenging. There should be $L^3 + 3L -1$ of them according to Eq. (\ref{eq:GSD-AB}), but it turns out not all of them are non-local. We will make a careful discussion of these operators, both local and non-local, in Sec. \ref{sec:logical-operator}.

\section{Logical operators}
\label{sec:logical-operator}

It is much easier to first think about the logical operators in the ABC model as there are only three of them, and they are quite easy to construct, as shown earlier in Fig. \ref{fig:mo}. This is the only kind of non-local operator that commutes with the ABC Hamiltonian. Using the identity $A_v = a_v A_v$ and the fact that a ground state is given as the projection $[\prod_v A_v ] (\otimes_l |0\rangle_l)$, one can show that a membrane operator acting on a ground state gives an identical state as another membrane operator, translated by one lattice spacing in the direction orthogonal to the membrane, acting on the same ground state. This explains why there are only three independent logical operators for the ABC model, and  ${\rm GSD} = 3^3$. Since all logical operators in the ABC model are non-local, one can say it has topological order.
%
%\begin{figure}[h]
%\centering
%\includegraphics[width=0.38\textwidth]{logical operator.png} 
%\caption{These two membrane operators act identically on the ground states of the ABC model. Red links mean the action of $x$ operator.}
%\label{fig:lo}
%\end{figure}

\begin{figure}[h]
\centering
\includegraphics[width=0.44\textwidth]{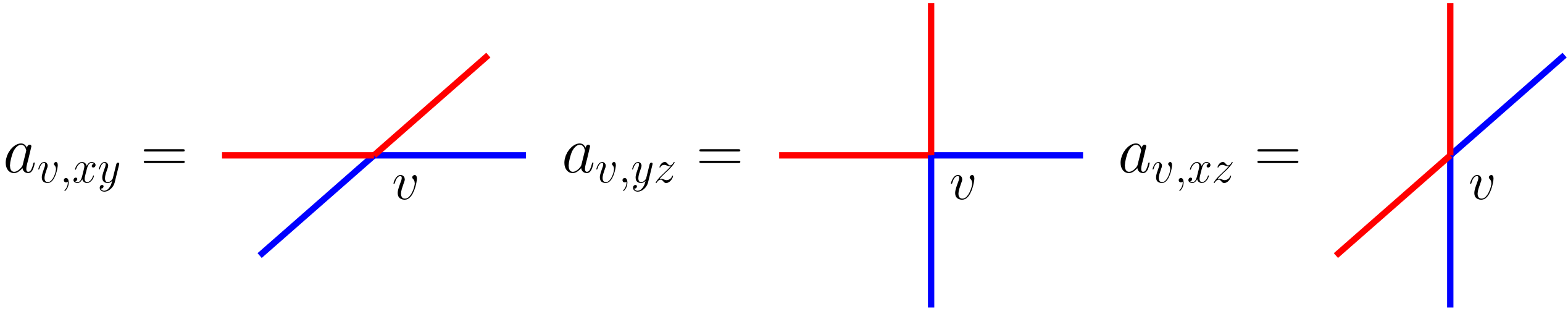} 
\caption{Vertex operators that serve as local logical operators of the AB model. Red and blue lines respectively represent $x$ and $x^2$.} 
\label{fig:a_v}
\end{figure}

For the AB model, we argued earlier that there will be $L^3+3L-1$ independent, logical operators. They can be classified as local and non-local. Here we are using the term ``logical operators'' as those that are capable of changing one ground state into another when acting on the initial ground state. The local logical operators are $a_{v,\lambda}$'s ($\lambda= xy,~ yz,~ xz$) shown in Fig. \ref{fig:a_v}, while the non-local ones are straight-loop operators in the usual sense. 

\begin{figure}[h]
\centering
\includegraphics[width=0.45\textwidth]{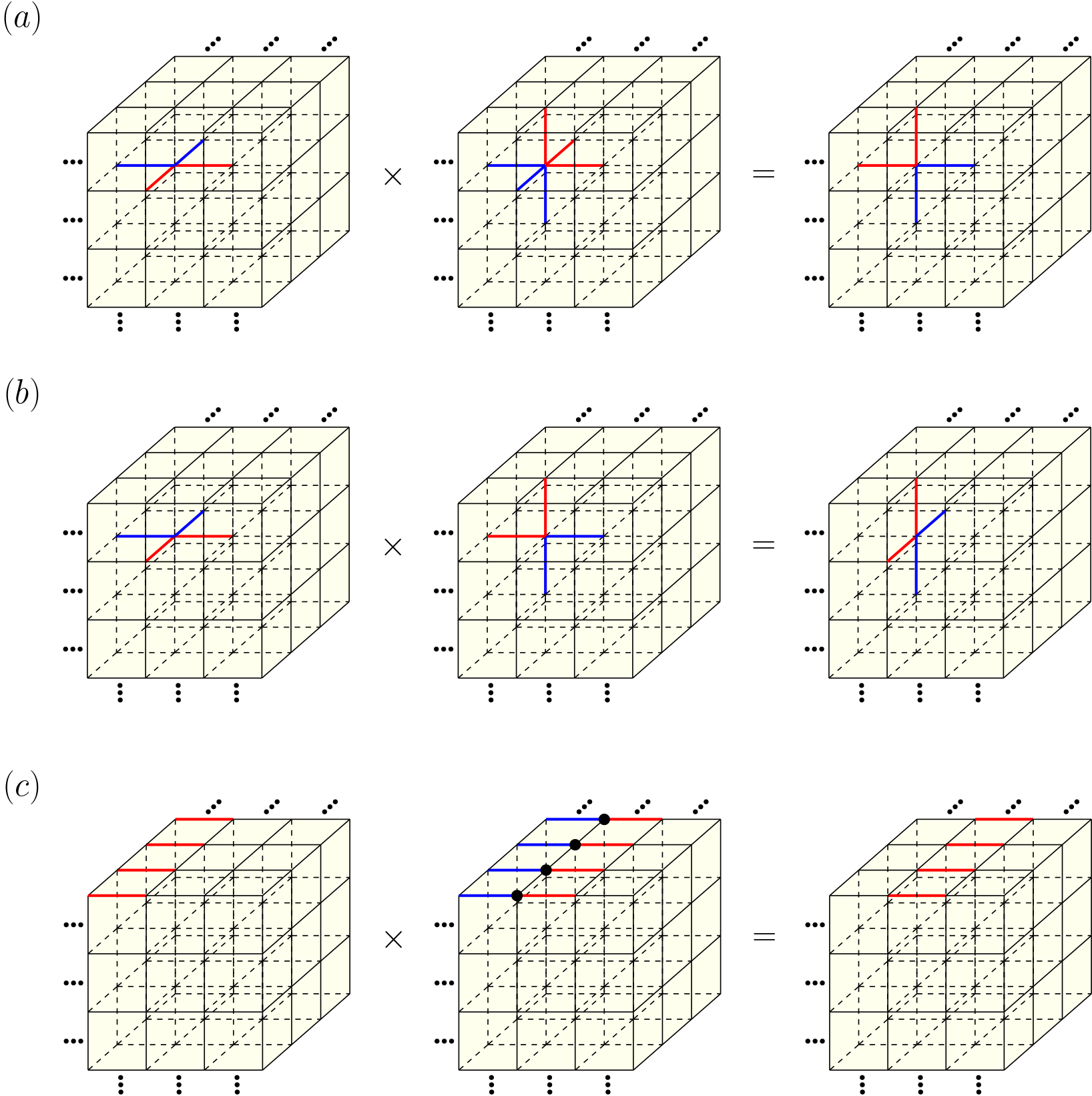} 
\caption{(a) $a_{v,yz}$ is the product of $a_{v,xy}^2$ on a given vertex and $a_{v}$. (b) $a_{v,xz}$ is the product of $a_{v,xy}^2$ on a given vertex and $a_{v,yz}$. (c) The product of $a_{v,xy}$ along the $x$ direction connects adjacent straight loop logical operators extended along the $x$ direction.}
\label{fig:a_v identity}
\end{figure}

We start with calculating the number of independent local logical operators.
First of all, the vertex operators $a_{v,\lambda}$ ($\lambda=xy,~yz,~xz$) shown in Fig. \ref{fig:a_v} commute with the existing $a_v$ and $b_c$ that define the vertex and cube operators of the AB model. These operators are, in fact, the $\mathbb{Z}_3$ versions of the vertex operators in the X-cube model, Eq. (\ref{eq:X-cube-A}). If we naively count the number of $a_{v,\lambda}$'s, there will be $3L^3$ of them, not all of which are independent. As shown in Fig. \ref{fig:a_v identity}(a),(b), one can see that $a_{v,yz}$ is equal to the product $a_{v,xy}^2 a_{v}$,  and $a_{v,xz}$ is equal to $a_{v,xy}^2 a_{v,yz}$. It then suffices to count the local logical operators that are made in terms of $a_{v,xy}$ only. The product of $a_{v,\lambda}$ in a given $\lambda$ layer is an identity as we described in Fig. \ref{fig:7}(a), implying that only $L^2-1$ $a_{v,xy}$ operators are independent in a given $xy$ layer. In the last $xy$ layer in a stack of $L$ layers of the cubic lattice, the counting argument applies differently as one should keep in mind not only the identity in the $xy$ layer but also the identities of the $yz$ and $xz$ layers, which gives rise to  $(L-1)^2$ independent local logical operators for the last $xy$ layer instead of $L^2 -1$. Therefore, the number of local logical operators is
\ba
(L^2-1)\times(L-1)+(L-1)^2=L^3-3L+2.
\ea

The rest of the logical operators are the straight loop logical operators given as the product $\prod_i x_i$ along a non-contractible straight loop in all three directions. One can easily check that they commute with all the vertex and the cube operators and that there are $6L-3$ of them which is also the number of logical operators in the X-cube model. In the toric code or the X-cube model, logical operators defined on adjacent straight lines are equivalent and do not produce new ground states when acting on a given ground state. In the AB model, they are {\it not equivalent}, but are still connected to each other by various local operators $a_{v,\lambda}$'s whose actions have already been accounted for. Figure \ref{fig:a_v identity}(c) shows how the two adjacent straight loop operators extended along the $x$ direction are connected by the product of $a_{v,xy}$'s. A similar argument applies to other orientations of straight loop operators. Although they are different non-local operators, they fail to produce any new ground states not already accounted for by the action of local operators. In summary, the $3^{L^3 + 3L - 1}$ ground states of the AB model are connected to one another by applying one of the $6L-3$ non-local operators or one of the $L^3 - 3L +2$ local logical operators. The non-local logical operators can be thought of as those of the $\mathbb{Z}_3$ X-cube model.

\section{Excitations and Braiding}
\label{sec:excitation}

The ABC model supports various excitations, dubbed fluxons, fractons, and freeons. They are the excitations taking place inside a plaquette, a cube, or at the vertex, respectively. The freeon is the three-dimensional analogue of the $e$ particle in the toric code and is free to move without any directional constraint. All the excitations in our model come in two colors (charges) due to having $\mathbb{Z}_3$ degrees of freedom at the links. It is helpful first to introduce some additional vertex, cube, and plaquette operators $A_v (n)$, $B_c (n)$ and $C_p (n)$, defined as
\ba
A_v (n)&=&\frac{1}{3}(1+\omega^n a_v+\omega^{2n} a_v^2)\nn
B_c (n)&=&\frac{1}{3}(1+\omega^n b_c+\omega^{2n} b_c^2)\nn
C_p (n)&=&\frac{1}{3}(1+\omega^n c_{p}+\omega^{2n}c_{p}^2)
\ea
where $n=0,~1,~2$. One can check the following relations
\ba
O(n)O(n')&=&\delta(n,n')O(n),\nn
\left[O(n),O'(n')\right]&=&0,\nn
\sum_n O(n)&=&1,
\ea
where $O=A_v,~B_c,~C_p$ and $O'=A_{v'},~B_{c'},~C_{p'}$. With these machinery at the ready, we discuss the fracton and fluxon excitations first, as they are closely related, and the freeon excitations later. 

\subsection{Fractons and Fluxons}

\begin{figure}[h]
\centering
\includegraphics[width=0.47\textwidth]{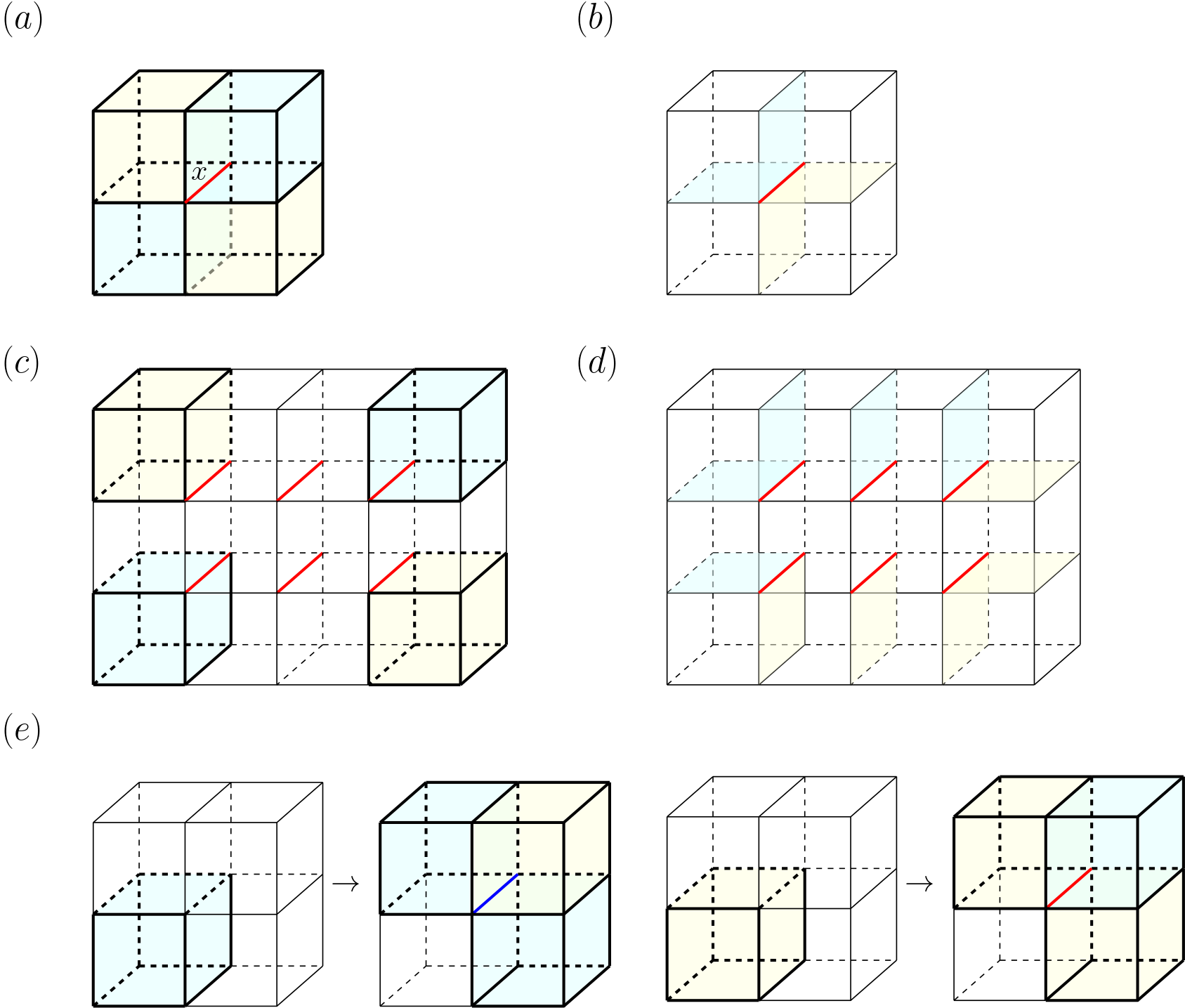} 
\caption{Applying a $x$ operator on a link creates (a) two blue fractons and two yellow fractons, as well as (b) two blue fluxons and two yellow fluxons in the ABC model. (c) Four fractons can be taken apart without extra cube energy, but (d)  costs plaquette energy growing as $\alpha$ times the number of fluxons. (e) A single blue or yellow fracton cannot move without creating two additional fractons.}
\label{fig:3}
\end{figure}

The fracton excitations in our model can be characterized with the help of $B_c(n)$. Acting on a ground state with $x$ on a single link ( a `defect link') as illustrated in Fig. \ref{fig:3}(a), eigenvalues of the four cube operators $B_c$ that share the defect link become $0$, leading to four fracton excitations on the adjoining cubes. These fractons are in turn distinguished in terms of their colors. Let us define $b_{c_b}$ and $b_{c_y}$ as the cube operators for the blue cube and the yellow cube, respectively, in Fig. \ref{fig:3}. One can show $b_{c_b} x=\omega x b_{c_b}$, $b_{c_y} x= \omega^2 x b_{c_y}$ by Eq. (\ref{eq:d3 relation}) and from this, it follows that
\ba
B_{c_b}(n) x\ket{G}&=&\frac{1}{3}(1+\omega^{n+1}+\omega^{2n+2})x\ket{G}\nn
B_{c_y}(n) x\ket{G}
&=&\frac{1}{3}(1+\omega^{n+2}+\omega^{2n+1})x\ket{G}.
\ea
Namely, blue cubes are the eigenstates of $B_c (1)$ with the eigenvalue $1$ while yellow cubes are the eigenstates of $B_c (2)$ with the eigenvalue $1$. Simultaneously, the eigenvalues of the plaquette operators $C_p$ for the four plaquettes sharing the defect link also change from $+1$ to $0$, costing energy $\alpha$ per plaquette as in Fig. \ref{fig:3}(b). Similar to fractons, fluxons are colored by blue or yellow as they become the eigenstates of $C_p(1)$ or $C_p(2)$ with eigenvalue $1$, respectively. These are the fluxon excitations. Starting from the four-fracton cluster in Fig. \ref{fig:3}(a), one can make continued insertions of the defect links as shown in Fig. \ref{fig:3}(c) without an extra cost in the cube energy. However, there are some plaquette excitations associated with each new defect link that cost energy $+\alpha$ each (see Fig. \ref{fig:3}(d)). Fractons and the accompanying fluxons are {\it confined} in the ABC model, in the sense that the expansion of the four-fracton cluster costs energy that grows as $\alpha$ times the number of accompanying fluxons. When we move a single blue or yellow fracton in any direction as in Fig. \ref{fig:3}(e), there is an extra blue-yellow cube pair left behind, as is characteristic of the fracton physics~\cite{vijay16}.

\subsection{Freeon Excitations}
Vertex excitations in the X-cube model are known as {\it lineons} as they are able to move freely (i.e. without extra energy cost) along only one direction. The vertex excitations in our ABC model, on the contrary, are able to move freely in any direction. Rather than being lineons, they behave like the vertex excitations in the two- and three-dimensional toric codes. 

\begin{figure}[h]
\centering
\includegraphics[width=0.4\textwidth]{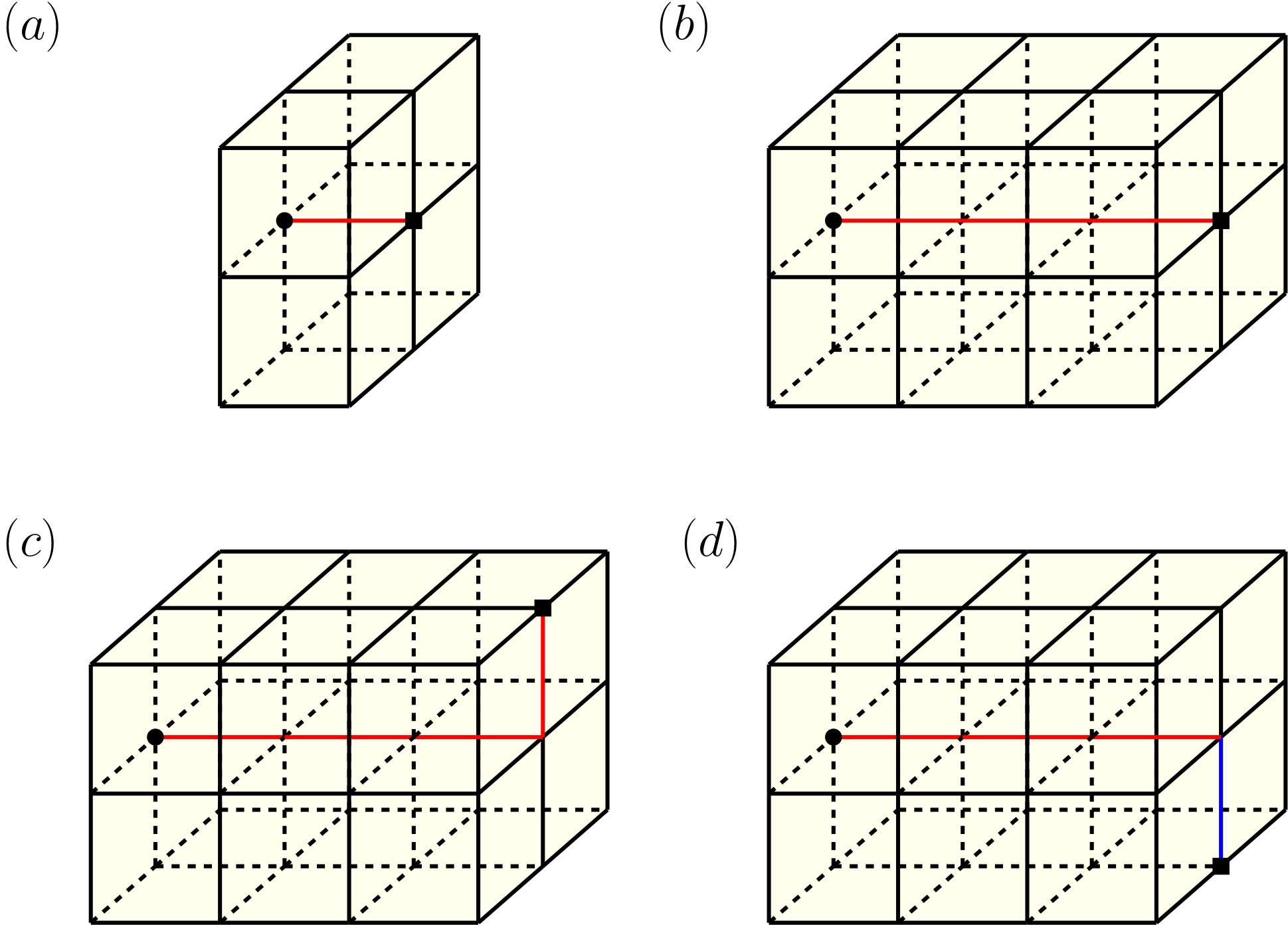} 
\caption{Red and blue lines respectively represent $z$ and $z^2$. (a) Applying a $z$ operator on a link creates two vertex excitations. (b) A square quasi-particle can move along the $y$ direction without an extra energy cost by continued application of $z$'s. A square quasi-particle can change its direction to (c) $-x$ or $+z$ direction without extra energy by acting $z$ and (d) $+x$ or $-z$ direction without extra energy by acting $z^2$.}
\label{fig:vertex-excitation}
\end{figure}

When we act $z$ on a link of a ground state as shown in Fig. \ref{fig:vertex-excitation}(a), eigenvalues of the two vertex operators $A_v$ whose vertices touch the link become 0, and two vertex excitations are created. In Fig. \ref{fig:vertex-excitation}, we marked them as $\CIRCLE$ (circle) and $\blacksquare$ (square) at the respective vertices. We can show
\ba
A_{v_{\scircle{}}}(n) z\ket{G}&=&\frac{1}{3}(1+\omega^{n+1}+\omega^{2n+2})z\ket{G}\nn
A_{v_{\blacksquare}}(n) z\ket{G}&=&\frac{1}{3}(1+\omega^{n+2}+\omega^{2n+1})z\ket{G},
\ea
where $A_{v_{\scircle{}}}(n)$ and $A_{v_{\blacksquare}}(n)$ are the vertex operators acting on the circle vertex and the square vertex, respectively. In other words, circle and square vertices are respectively the eigenstates of $A_v (1)$ and $A_v (2)$ with the eigenvalue $1$, implying two different charges of the freeon excitation.

To separate the freeon pair, one keeps applying the $z$ operator along a continued line segment as in Fig. \ref{fig:vertex-excitation}(b). To create a turn, one applies either a $z$ or $z^2$ operator at the link orthogonal to the original line segment as shown in Fig. \ref{fig:vertex-excitation}(c) or (d). The choice is made in such a way that the freeon pair configuration commutes with all the vertex operators $A_v$ except the two at the ends. As a result, vertex excitations can move in any direction without having to further create residual vertex excitations. The scheme fails in the case of X-cube or its $\mathbb{Z}_n$ generalizations due to the fact that there are always three kinds of vertex operators $A^{xy}_v,~ A^{yz}_v,~ A^{xz}_v$, and a turn in the path of the string operator is bound to create excitations in at least one of them. There is only one vertex operator in the ABC model (as in the three-dimensional toric code), and finding a freeon path that commutes with the vertex operators becomes possible. The co-existence of vertex excitations with no directional restriction and the fracton excitation with restricted mobility places our model in a distinct category from either Type I or Type II fracton models.

\subsection{Braiding}
Thanks to the unrestricted mobility of the freeon, one can imagine an adiabatic motion of a freeon and a non-trivial phase picked up in the process. Since the freeons are the only excitations with truly unrestricted mobility, it is natural to think of a freeon trajectory in the background of other excitations that are held fixed. 

\begin{figure}[h]
\centering
\includegraphics[width=0.44\textwidth]{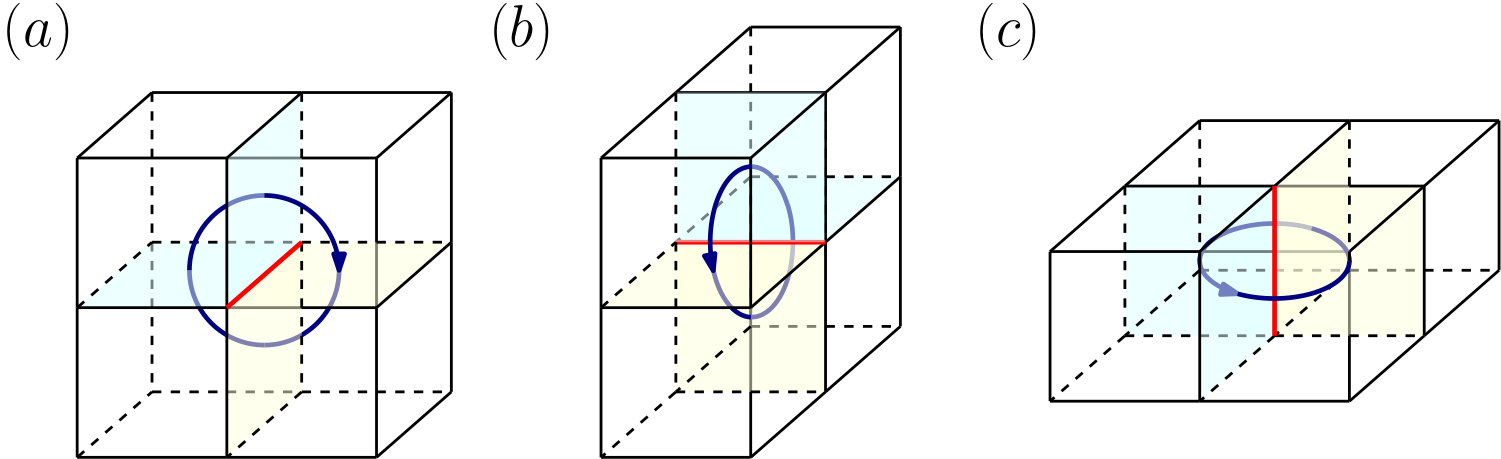} 
\caption{Magnetic flux lines emanating from the four-fluxon cluster form a closed ring of effective magnetic flux. Red lines represent $x$. Blue and yellow plaquettes represent $\omega$ and $\omega^2$ plaquette quantum numbers (fluxes).}
\label{fig:boundary}
\end{figure}

\begin{figure}[h]
\centering
\includegraphics[width=0.40\textwidth]{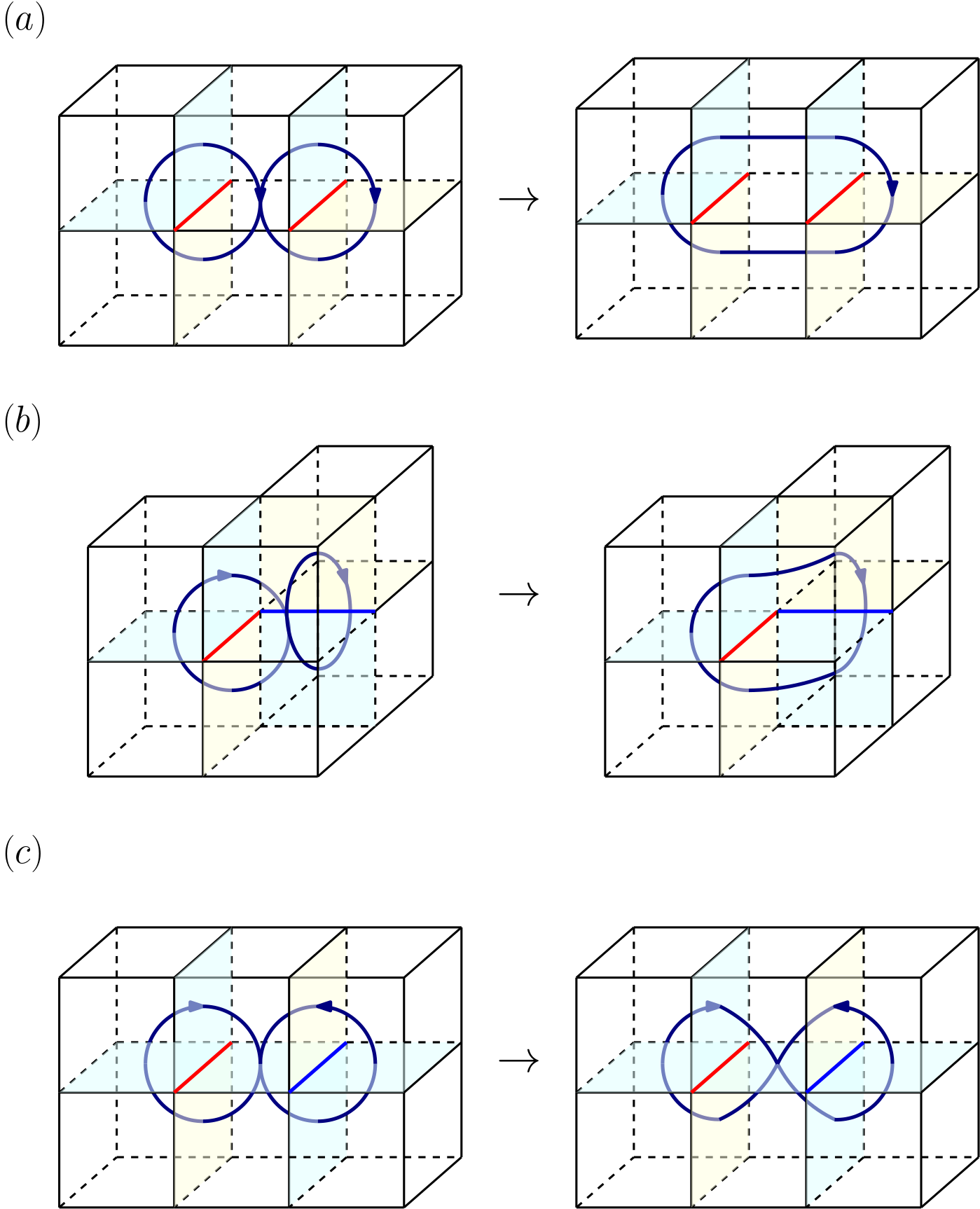} 
\caption{(a) The fluxon-enclosing path can be enlarged as more fluxons appear. (b) The fluxon-enclosing path can make a turn as the fluxons appear on different planar orientations. (c) A figure-eight path is associated with this configuration of fluxons.}
\label{fig:boundary2}
\end{figure}

Although fractons and fluxons tend to be created in tandem, for ease of illustration we display only the fluxon clusters in Fig. \ref{fig:boundary}. Each fluxon emanates a ``magnetic flux" in the direction perpendicular to the plaquette. Depending on the color of the fluxon, one can assign the direction of the magnetic flux to the plaquette. By smoothly connecting the fluxes emanating from the plaquettes, one arrives at a closed path shown as circles in Fig. \ref{fig:boundary}. The path is directed, pointing at the ``positive" direction of the magnetic flux. The fluxon boundaries can be deformed in various ways. Figure  \ref{fig:boundary2}(a) depicts the situation where, as the fluxon cluster expands from having four to six fluxons, the encircling path expands along with it. The fluxon loop does not have to be confined to a plane, as shown in Fig. \ref{fig:boundary2}(b). It can even be a figure-eight shape, as in Fig. \ref{fig:boundary2}(c), implying that a $x$ and a $x^2$ link excitation will not merge and annihilate easily.

\begin{figure}[h]
\centering
\includegraphics[width=0.44\textwidth]{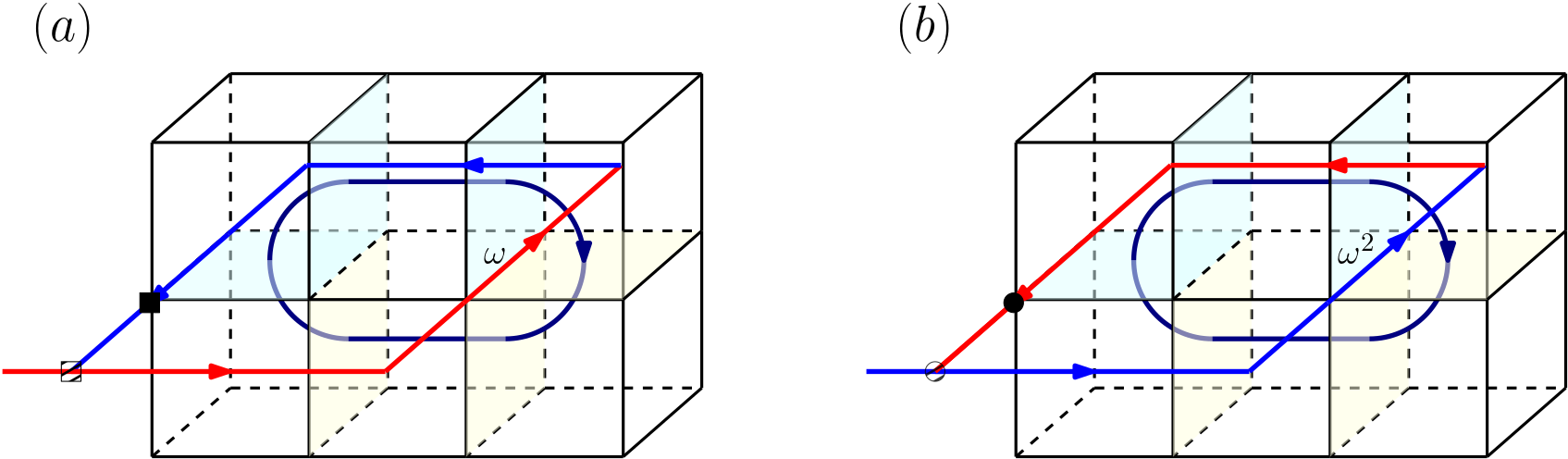}
\caption{Red and blue lines represent $z$ and $z^2$, respectively. (a) When a square freeon enter the fluxon loop in a clockwise fashion and move back to the original position, it gives a phase factor $\omega$. (b) When a circle freeon enter the fluxon loop in a clockwise fashion and move back to the original position, it gives a phase factor $\omega^2$.}
\label{fig:braiding}
\end{figure}

Now imagine a freeon of either charge (a circle or a square freeon) entering the fluxon loop in a clockwise fashion, i.e. seeing the arrows in the fluxon loop as going clockwise as the freeon enters the region enclosed by it. The freeon path must intersect one of the fluxon defect links made by either $x$ or $x^2$ and pick up a phase of $\omega$ or $\omega^2$ as it moves back to its original position (Fig. \ref{fig:braiding}). For the entrance into the counterclockwise fluxon loop, the phase factor will reverse. The closed loop made by a fluxon cluster can be viewed as the loop of magnetic flux or a vortex loop. The analogy becomes natural in $\mathbb{Z}_3$ models as the two fluxon charges can be viewed as directions of the magnetic flux. There is no sense of direction for the $\pi$-flux excitations in the $\mathbb{Z}_2$ models. One can say the $m$ particles of the $\mathbb{Z}_3$ toric code are now created three-dimensionally, forming the flux loops. The charged particles - $e$ particles in the toric code and freeons in our model - pick up the Aharonov-Bohm phase by going round the flux loop. The mutual statistics of $e$ and $m$ particles in the toric code can be interpreted as the effective magnetic flux of $\pi$ carried by the other species. Here in our model, the effective flux of $2\pi/3$ is concentrated along the fluxon loop. A freeon moving adiabatically around a closed path can detect the presence of fluxons through the phase factor it picks up during the passage.

\begin{figure}[h]
\centering
\includegraphics[width=0.44\textwidth]{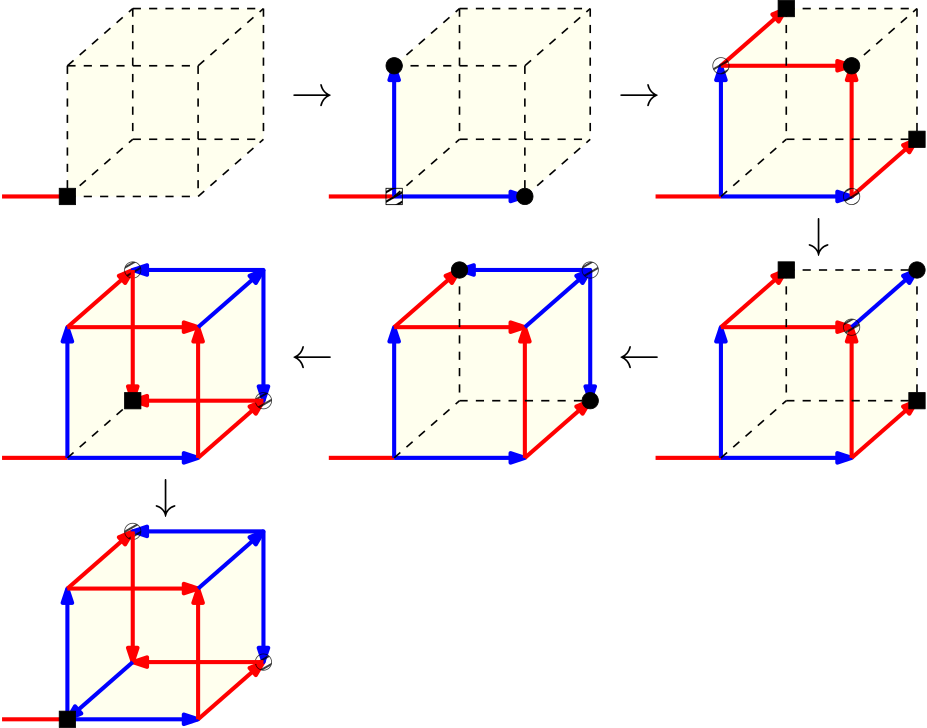} 
\caption{Schematic figure of the procedure of square freeon braiding in the AB model. Red lines and blue lines represent $z$ and $z^2$, respectively. In each step, newly created circle or square freeons are fully filled and annihilated ones are scarcely filled. }
\label{fig:braiding2}
\end{figure}

One may ask: is it possible to use freeons to detect the presence of fractons as well? The answer is yes, as already well explained in recent papers~\cite{hermele19a, hermele17}. We give an adaptation of the existing argument that suits our model. This time, the freeon path is defined utilizing the $b_c$ operators (not $z$ operators), and moving a freeon by a series of multiplication of $b_c$ or $b_c^2$ operators for arbitrary size of the cube can be understood by the fusion rule of freeons. When we have a square freeon, for instance, if a square freeon move along the edges of the arbitrary size of the cube, we can initiate the statistical interaction of the freeon and a fracton by first separating the square freeon into two circle freeons and moving them to the vertices that are placed at $-y$ and $+z$ directions from the origial vertex (the second diagram of Fig. \ref{fig:braiding2}). After that, two circle freeons change to one circle freeon plus two square freeons, and the three of them move to the vertices marked in the third diagram of Fig. \ref{fig:braiding2}. Following through the procedures outlined in Fig. \ref{fig:braiding2} where each step relies on the fusion rule of freeons, the original square freeon state is restored, but with multiplication by the operator $b_c^2$. Therefore, in effect, the process will have measured the presence of a fracton regardless of which kind of fluxon distribution accompanies the fracton creation. When we have a circle freeon instead, the procedure described in Fig. \ref{fig:braiding2} is equal to multiplication by $b_c$ instead. In either case, the procedure results in a non-trivial phase factor if there is a fracton inside the cube. These remote detection methods for fractons were introduced in Refs.~\onlinecite{hermele19a, hermele17}. We have outlined in Fig. \ref{fig:braiding2} the $\mathbb{Z}_3$ version of the remote detection scheme for fractons, applicable for both AB and ABC models.

\begin{figure}[h]
\centering
\includegraphics[width=0.40\textwidth]{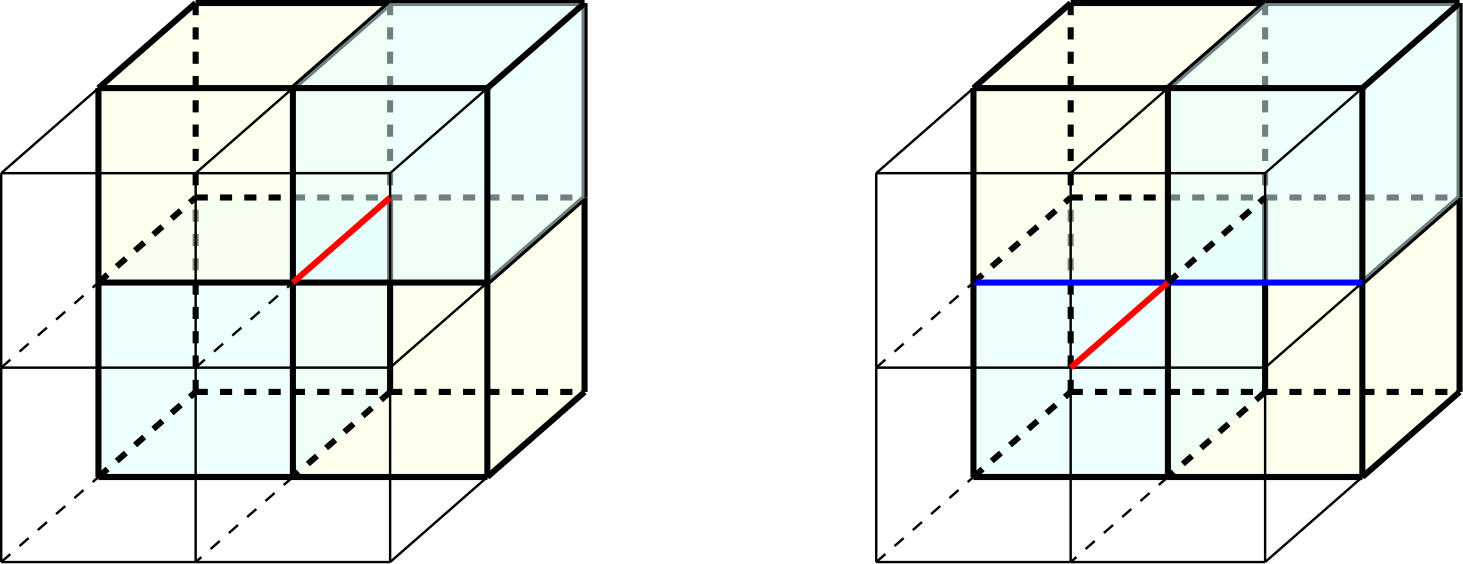}
\caption{Constructing an identical fracton cluster in two different ways. Red lines and blue lines represent $x$ and $x^2$, respectively. Note that their fluxon contents differ.}
\label{fig:fracton cluster}
\end{figure}

The reason why the above elaborate detection scheme for fractons is essential is that the freeon loop used to detect the fluxon cluster fails to give a unique answer when it comes to detecting fractons. To illustrate why, Fig. \ref{fig:fracton cluster} shows two identical fracton clusters that differ in their fluxon contents. A freeon path such as given before obviously picks up different phase factors in the two situations, although the fracton contents are the same in both. Note that the two operations shown in Fig. \ref{fig:fracton cluster} are related by $a_{v,xy}$. Meanwhile, a closed freeon path like those in Fig. \ref{fig:braiding} is generated by the product of $c_{p,xy}$'s and $c_{p,xy}^2$'s, which do not commute with $a_{v,xy}$. This is why the two fracton configurations in Fig. \ref{fig:braiding} give rise to different freeon phases.

To sum up, the detection of both fluxons and fractons by the adiabatic evolution of freeons can be done in the ABC model but only the detection of fractons is meaningful in the AB model. The ABC model and the $\mathbb{Z}_3$ toric code in three dimensions share the same GSD, but the big difference arises in the existence of fractons in the ABC model but not in the toric code. As a result, freeons experience phase factors through statistical interaction with fractons as well as the fluxons, while only the fluxons are responsible for the adiabatic phase of freeons in the toric code. 

\section{Discussion}

We have presented a new kind of fracton model distinct from previous models in (i) the existence of local symmetries and (closely related) extensive GSD, (ii) the existence of both non-local and local logical operators connecting different ground states, and (iii) free vertex excitations called freeons with non-trivial mutual statistics with respect to the fracton-fluxon excitations. There are mutually commuting the vertex ($A_v$), cube ($B_c$), and plaquette ($C_p$) terms in the model. The extensive GSD is present only in the AB model with the vertex and the cube terms present. The logical operators that help alter one ground state into another have been sorted out for both AB and ABC models. The fracton excitations are accompanied by the fluxons, and the latter objects tend to create a linear potential between the fractons leading to the confinement of both in the ABC model. The vertex excitations called freeons, on the other hand, remain completely free to move in any direction both in the AB and ABC models. The freeon and the local plaquette excitation exhibit mutual anyonic statistics that can be detected whenever the freeon path crosses the loop encompassing the fluxon cluster. 

Past generalizations of the X-cube model involved geometric deformations of the cubic lattice~\cite{slagle18, chen18} and/or enhancing $\mathbb{Z}_2$ to $\mathbb{Z}_N$ degrees of freedom at the links~\cite{vijay17, slagle17, slagle18, chen18}. Common to these models is the existence of three kinds of vertex operators at the vertex, one for each planar orientation. This is one route to generalize the vertex terms in the two-dimensional $\mathbb{Z}_N$ toric code, of course, but our proposal here defines another route at the generalization. The key idea is the introduction of only one vertex operator $A_v$ consisting of $x^{-1}$ for half the links and $x$ for the other half of the links connected to a vertex. Such construction works well for $\mathbb{Z}_3$ link variables but fails to produce a vertex operator that commutes with the cube operator if the local Hilbert space is $\mathbb{Z}_2$. In other words, $\mathbb{Z}_3$ local Hilbert space is essential for our construction to work. Other properties of the model such as the orientation of the flux and flux loop excitations also derive from $\mathbb{Z}_3$ and absent in $\mathbb{Z}_2$ models. 

It turns out the three-dimensional $\mathbb{Z}_3$ toric code assumes exactly the same kind of vertex operator as ours but, instead of a cube operator, has three sets of plaquette (or flux) operators~\cite{wen05}. Our model in the absence of the cube term, i.e. the AC model, is in fact the three-dimensional $\mathbb{Z}_3$ toric code. The vertex excitations in the toric codes are also free - a property that our model inherits despite also having fracton-like excitations. In many respects, our model is a hybrid between the X-cube and the three-dimensional toric code and realizes properties of both, most notably the fracton excitations and the free vertex excitations. 

Our model study suggests that the sub-extensive GSD is not a necessary ingredient for realizing fracton behavior. It will be interesting to see how the higher-rank gauge theory formulation of the fracton dynamics first suggested in Ref. \onlinecite{pretko17} will play out in our model. Characteristics of fluxon and freeon excitations we analyzed in the ABC model might also lead to robust error correcting code whose nature is akin to that of the 3D toric code. In the 3D toric code, different ground states that retain the quantum bit of information are connected by the membrane operators whose energy costs due to the fluxon creation are $O(L^{2/3})$, which makes it a very stable quantum memory~\cite{michnicki14}. The mechanics of the ABC model is quite similar to 3D toric code and the model may well serve as an error correcting code.  Since the creation of fractons accompanying the fluxons further adds to the energy cost, one might expect even more stability as a quantum memory from the ABC model.

%Physically, this is accomplished by creating an anyon pair (at finite energy cost) and moving them around the entire circumference of the torus for re-annihilation. %Instead of $\mathbb{Z}_2$ toric code, we can imagine a $\mathbb{Z}_2$ version of AC model having the vertex and the plaquette operators only. The GSD of such model will be $2^3$, encoding three qubits. In this model, different ground states are only connected by the action of a membrane operator rather than a loop operator. One can easily estimate the energy cost of making a membrane operator to be of order $O(L^2)$, much larger than the $O(1)$ energy cost of an anyon pair in the toric code and hence much more difficult to accomplish physically. Protection by membrane is far more stable than the one by a loop, and one might think of the $\mathbb{Z}_2$ ABC model as a more stable error correcting code than the toric code. 

\acknowledgments H. J. H. was supported by the Quantum Computing Development Program (No. 2019M3E4A1080227).  

\bibliography{SC}
\end{document}